\documentclass[journal]{vgtc}                     

\onlineid{1438}

\vgtccategory{Research}

\vgtcpapertype{Analytics \& Decisions}

\title{Defogger: A Visual Analysis Approach for Data  Exploration\\of Sensitive Data Protected by Differential Privacy}

\author{%
  Xumeng Wang,
  Shuangcheng Jiao, and 
  Chris Bryan
}

\authorfooter{
  \item
  	Xumeng Wang and Shuangcheng Jiao are with DISSec, Nankai University.
  	E-mail: wangxumeng@nankai.edu.cn, jiaoshuangcheng@mail.nankai.edu.cn.

  \item Chris Bryan is with SCAI, Arizona State University.
  	E-mail: cbryan16@asu.edu.
  \item Xumeng Wang is the corresponding author.
}

\newcommand{\highlight}[1]{{\color{black}#1}}
\newcommand{\techname}{Defogger}

\abstract{%
  Differential privacy ensures the security of individual privacy but poses challenges to data exploration processes because the limited privacy budget incapacitates the flexibility of exploration and the noisy feedback of data requests leads to confusing uncertainty. In this study, we take the lead in describing corresponding exploration scenarios, including underlying requirements and available exploration strategies. To facilitate practical applications, we propose a visual analysis approach to the formulation of exploration strategies. Our approach applies a reinforcement learning model to provide diverse suggestions for exploration strategies according to the exploration intent of users. A novel visual design for representing uncertainty in correlation patterns is integrated into our prototype system to support the proposed approach. Finally, we implemented a user study and two case studies. The results of these studies verified that our approach can help develop strategies that satisfy the exploration intent of users.
}

\keywords{Differential privacy, Visual data analysis, Data exploration, Visualization for uncertainty illustration}

\teaser{
  \centering
  \includegraphics[width=\linewidth]{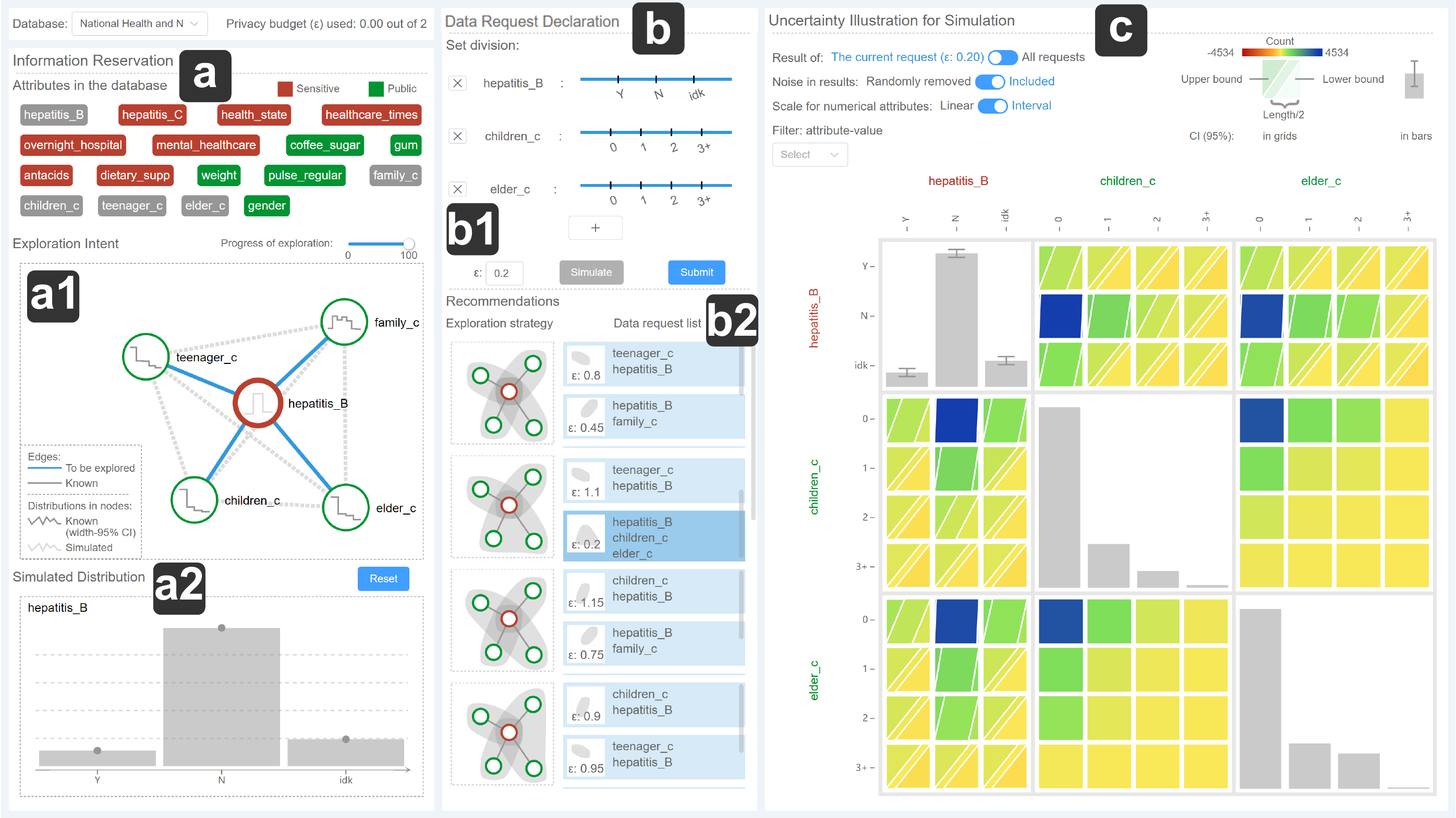}
  \caption{The interface of \techname{} for exploring data guarded by differential privacy. (a) The information reservation view invites users to specify data information (e.g., distributions and correlations) of interest as their exploration intent (a1) and describe available data facts to generate simulated data (a2) for exploration strategy recommendation. (b) The data request \highlight{declaration view} allows users to declare a data request (b1) based on recommendations from a reinforcement learning model (b2). (c) The uncertainty illustration view explains the uncertainty caused by differential privacy in the simulated or the actual response of the declared data request.
  }
  \label{fig:teaser}
}

\graphicspath{{figs/}{figures/}{pictures/}{images/}{./}} 

\usepackage{tabu}                      
\usepackage{booktabs}                  
\usepackage{soul}
\usepackage{lipsum}                    
\usepackage{mwe}                       
\usepackage{CJKutf8}                   
\usepackage{colortbl}
\usepackage{mathptmx}                  

\begin{document}
\begin{CJK}{UTF8}{gbsn}

\firstsection{Introduction}
\maketitle

As datasets that contain sensitive information, such as medical, social, or other personal data~\cite{kwon2018retainvis, ouyang2023leveraging, shi2020urbanmotion}, become increasingly prevalent, the querying, \highlight{browsing, analyzing, and exploring} of such data must enforce privacy-related restrictions. When providing data based on a user request (e.g., as part of a database query or API call), a common mechanism to enforce privacy restrictions is the use of \textit{differential privacy} \highlight{(DP)}~\cite{dwork2014algorithmic}, which protects data records by injecting noise into the \highlight{response to data requests} (e.g., as part of a data exploration process~\cite{battle2019characterizing}). \highlight{DP} has become a common mechanism to scaffold access for databases that contain sensitive information because they have been shown to be effective in safeguarding privacy during exploration workflows~\cite{li2022dplanner} while also meeting regulatory requirements such as the European Union’s GDPR~\cite{voigt2017eu} and California’s CPRA~\cite{bukaty2021california}. 

\highlight{DP} approaches commonly maintain a \textit{privacy budget}: when a data request is issued, a spending deduction is calculated based on the amount of noise added to the response and this is subtracted from the overall budget amount~\cite{ebadi2015differential}. \highlight{A higher budget can be traded for a response with a smaller noise.} Importantly, once a response is provided, there is no mechanism for rolling back the request or increasing the budget. As the budget is exhausted, it becomes difficult to gain insights because responses will be highly noisy, which leads to high uncertainty. \highlight{Therefore, DP is strained for data exploration processes.} In particular, \textit{exploratory visual analysis}, whereby humans create and inspect a series of visualizations to explore, analyze, or discover insights from the dataset~\cite{battle2019characterizing}, suffers more from \highlight{DP} approaches: \highlight{(i) The noise injected by DP can mask the visualized patterns (e.g., trends, outliers)~\cite{zhang2020investigating} or be overpriced~\cite{wang2021current}. (ii) A user will likely explore the dataset by browsing a series of charts~\cite{kwon2018retainvis, ouyang2023leveraging, shi2020urbanmotion}, while each chart must be drawn based on requested results (e.g., results of count queries for drawing a histogram), thus continually consuming a given privacy budget. (iii)~Consumption is hard to predict because users could send requests based on serendipitous findings from a previously-created chart, or perform ``locate'' or ``explore'' search actions~\cite{brehmer2013multi}, which makes a \textit{priori} customized noising solutions (such as those invoked by differential private data synthesis~\cite{zhang2021privsyn}) difficult to design and implement.}

The takeaway is that users need intelligent strategies for visually exploring datasets containing sensitive information that is protected by \highlight{DP}, as a naive approach can easily exhaust a privacy budget and result in uninformative charts that contain high uncertainty~\cite{zhang2020investigating}. Bolstered by a \highlight{use case that} identified an explicit set of context requirements, we thus propose a novel, first-of-its-kind workflow for this process, motivated by the need for users (i) to maintain sufficient awareness of budget allocation costs and spending, balanced against (ii) the need to support sufficient analysis and interrogation capabilities to derive insights and value from the dataset, while also (iii) adhering to required data safeguards imposed by \highlight{DP}. 

A key to our approach is that the pipeline considers that budget allocation is not only numerical calculations \highlight{of the noise size} but also a question of which responses actually merit spending the budget. Our solution is to provide recommendations modeled on a combination of the exploration interests of users, knowledge of data patterns, and an understanding of how a group of data requests can contribute to the exploration process. In part, we are inspired by prior tools (for non-sensitive datasets) that recommend `next steps' to augment visualization-driven exploration, such as Voyager~\cite{wongsuphasawat2017voyager}. Our pipeline goes a step further: we extract data patterns at different granularities (from overview to details) and leverage these to suggest responses tailored to an ``appropriate'' level of detail for visualizations that still successfully support exploration \highlight{but spend budget more wisely.}
We implement our pipeline in an interactive tool called \textit{Defogger}, which interactively \highlight{recommends feasible exploration strategies for the next step by a reinforcement learning model and demonstrates how each strategy will work (i.e., how much budget it will spend, and how much uncertainty the resultant visualization will contain) before actually applying the step.} 

We validate our pipeline (and \techname’s implementation) by conducting robust human-centered evaluations (a user study and case studies). Evaluation results indicate that our approach can improve returns on investment of privacy budgets through the intelligent recommendation of data requests based on the exploration intent of users. 
\textbf{\textit{Ultimately, our pipeline represents a novel approach that augments the ability of humans to explore and gain increased value from data while adhering to \highlight{DP} constraints.}} We conclude this paper by additionally contributing a discussion of generalizable lessons learned about the effective intersection of computational models, visualization, and interaction as a way to optimize privacy preservation and human sensemaking, which can benefit both the visualization and privacy communities, as well as limitations of the current pipeline and \techname{} designs, and suggest opportunities for future work in this area.

\section{Related Work}


\subsection{Processes of Exploratory Visual Analysis}
Exploratory visual analysis \highlight{of abstract/non-spatial data} mainly leverages three categories of operations, consisting of ``query,'' ``browse,'' and ``search'' actions to \highlight{iteratively interact with data}~\cite{heer2012interactive, battle2019characterizing}. If information needs to be excluded from the exploration target because of privacy concerns, query operations are necessary for requesting visualizations that can enable browsing and searching patterns of groups. Queries in the exploration process can be organized either in a top-down manner~\cite{zgraggen2018investigating} following a specific exploration goal or in a bottom-up fashion~\cite{alspaugh2018futzing} without a clear destination~\cite{battle2019characterizing}. When implementing a bottom-up exploration process, users have no predetermined target of visualization and therefore need recommendations for the next visualization. As for a top-down exploration process, recommendations help inspect whether there are ignored data patterns.

Recommendations for the next visualization can be made according to the history of users' exploration process. For example, prior tools~\cite{wongsuphasawat2017voyager, zhao2020chartseer, battle2016dynamic, wongsuphasawat2015voyager, qian2021learning} have considered the relevance of previously explored visualizations based on similarities in their data content or \highlight{data patterns.} Zhou et al.~\cite{zhou2021modeling} monitored the development of the exploration focus (based on which attributes were being explored) and developed a model to infer the next focus. \highlight{Qian et al.~\cite{qian2021learning} recommend visualizations based on the data patterns that could be of interest to the users.} Importantly, \textbf{\textit{none of the above studies consider privacy issues during exploration and recommendation}}. When constrained by privacy budgets, users \highlight{are unable to leverage a pattern-driven exploration process. Instead, they} have to carefully plan what content they should focus on during their exploration process. This paper addresses this gap by developing a recommendation pipeline that understands vague descriptions of users' interests in data content and makes corresponding recommendations. 

\subsection{Practices in Differentially Private Publication}
To apply \highlight{DP}, users have to develop and follow ``trading schemes,'' consisting of the group of trading targets and how much of the privacy budget should be spent on each target. Existing studies proposed examples of trading schemes and provided tools for scheme assessment.

In real-world scenarios, the values of data distribution in different bins are not the same. Users prefer to spend a higher privacy budget on significant bins than others. As the significance of each bin is unknown, trading rules can allow users to pay an extra privacy budget for significance assessment. For example, publishing differentially private histograms could benefit from an assessment result of histogram structure~\cite{zhu2017differentially, xu2013differentially}. When exploring data progressively, the error of prior publications can affect the significance of the next. An adaptive trading scheme~\cite{li2015differentially} pays a high budget for the next to avoid going astray when the cumulative error of prior publications is excessive. 

For the assessment of trading schemes, the accuracy of potential publication is a key metric. The accuracy derived from a trading scheme is not static due to the random noise. As the publication is a single value, visualizations~\cite{john2021decision, nanayakkara2022visualizing} can directly represent the distribution of potential publication results to summarize the effectiveness of a user-specified privacy budget. If the publication is a complicated pattern, the accuracy can be assessed by comparing the differentially private publications with the original ones~\cite{zhou2022dpviscreator, wang2017utility}.

Accessing the original data is necessary for existing studies to determine trading schemes because they consider the original data as input. The target users in this study are not authorized to access the original data. Therefore, determining and assessing trading schemes requires tolerating more uncertainty.
    
\subsection{Visualizations for Uncertainty Representation}

When there is uncertainty in data, understanding uncertainty is necessary for data analysis~\cite{hullman2018pursuit}. Visualizations can facilitate the understanding of uncertainty by demonstrating the existence of uncertainty in data distribution and representing the features of uncertainty. 

Indicators, like the level of confidence, quantify the degree of uncertainty. Classic visualizations (e.g., histograms, pie charts, and scatter matrices) that are designed to show distributions of 1-dimensional data or multi-dimensional data require extra visual representations to illustrate the degree of uncertainty simultaneously. Extra representations for uncertainty should not conflict with those used to visualize data distributions. Color encoding for uncertainty can be superimposed on visualizations that show data distributions by shape-related encodings or position encodings~\cite{sanyal2009user}. If the encoding of color filling is occupied, the style of strokes (e.g., opacity~\cite{zhang2021conceptscope}, wave frequency~\cite{gortler2017bubble}) can be customized for uncertainty representation. 

In addition to the degree, uncertainty analysis may require more detailed descriptions of uncertainty. To support uncertainty analysis, values with uncertainty can be depicted as a probability distribution and instantiated as a group of values. Kay et al.~\cite{kay2023ggdist} summarized four categories of visualizations for instantiating data with uncertainty, consisting of intervals, ribbons, slabs, and dotplots. These 4 categories can adapt to different visual representations of values with uncertainty or 1-dimensional data with uncertainty. Visualizition of uncertainty has previously been employed as a way to support privacy preserving visualizations and \highlight{DP}~\cite{nanayakkara2022visualizing, john2021decision, wang2017utility}. However, prior work does not consider data exploration based on a user's privacy budget, as is done in this paper. Also, users may explore complicated data patterns, like the distributions of multi-dimensional data. Nevertheless, uncertainty in distributions of multi-dimensional data still lacks effective representations. We propose a Mosaic design that can embed visual representations for uncertainty into the grid of heatmap matrices.

\section{Scenario Overview and Requirements Analysis}
Our target application scenario is centered around users who need to explore \highlight{tabular datasets containing sensitive information which are protected by DP mechanisms.} 
A real-world example of this would be the Cloud Healthcare API provided by BigQuery~\footnote{\url{https://cloud.google.com/bigquery/docs/differential-privacy}}. Users (generally researchers or clinicians) apply for portal access; once approved, they are allocated an exploration budget. The primary research question we investigate is, \textbf{\textit{how do we optimize the visual exploration process (thus providing value to these users) while maintaining appropriate privacy safeguards for sensitive data?}}
In this section, we briefly formalize the ``privacy aware exploration'' practice conducted by these sorts of users, outline a virtual use case to illustrate available exploration schemes and user requirements, and distill a set of requirements.


\subsection{Practice under Privacy Limitations}
\highlight{DP} forces users to obtain noisy responses to data requests based on a defined privacy budget. 
Formally, the mechanism of noise addition aims at yielding indistinguishable responses for two datasets $D$ and $D'$ differing by any single record.
Specifically, a data request $K$ with a privacy budget of $\varepsilon$ satisfies
$$Pr[K(D)\in S]\leq \exp(\varepsilon)\times Pr[K(D')\in S],$$
\highlight{where $S$ could be any subset of all possible values that can be returned by $K$. In other words, $K$} needs to yield similar responses for the two datasets. Hence, the noisy responses can prevent sensitive information about an individual from leaking. A lower privacy budget denotes a higher limitation of privacy preservation, which requires a relatively larger noise. To ensure that the size of the noise to be added is appropriate, mechanisms of noise generation consider $\Delta f$, the sensitivity of responses, which is how much difference could exist between the actual response on $D$ and $D'$. For example, the count queries used to summarize data distribution have a unified sensitivity of 1. The Laplace mechanism yields random noise with a Laplace distribution $Lap(\Delta f/ \varepsilon)$. According to the Laplace mechanism, the results of count queries will be added random noise satisfying $Lap(1/ \varepsilon)$. \highlight{Since the size of noise is irrelevant to the value of query results, the disturbance (or uncertainty) caused by noise is relatively small to a large value.}

\begin{figure}[h]
    \centering
    \includegraphics[width=0.8\columnwidth]{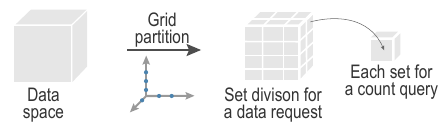}
    \caption{\highlight{Set divisions used in a data request.}}
    \label{fig:sd}
\end{figure}

\highlight{In practice, users are not permitted to have direct access to individual information. Instead, they can request statistical information for groups. The charge of privacy budget for statistical information is independent of the group size because data requests against other individuals will cause no additive risks to privacy exposure~\cite{li2019personalized}.} Users can determine the cost of the privacy budget for a statistical request only based on the expected size of the noise. \highlight{In contrast, repeatedly requesting the same individual's information increases the risk of identity or information leakage. In this case, the privacy budget must be charged again~\cite{xu2013differentially}. For instance, count queries (i.e., quantity statistics) for multiple sets can be executed in a single data request and charged once when there is no intersection among all sets because each individual needs to be counted in at most one set. Following this principle, users can optimize the use of a privacy budget by employing multiple sets, which can cover the entire dataset without any overlap. To identify such set groups for data tables, users can declare conditions (i.e., intervals for numerical attributes and categories for categorical attributes) for \textit{set division} according to a grid partition of the data space (see Fig.~\ref{fig:sd}).}

\highlight{Data requests can employ either a fine or a coarse granularity of grid partition to determine a set division for batch queries. A fine granularity of set division can facilitate flexible exploration of data distribution from different perspectives because the total count of a group of sets can be considered the count of their parent set. }
Nevertheless, the sum calculation integrates the noise added to each count of small sets together. Assume that a privacy budget $\varepsilon$ is paid for \highlight{a data request}. The noise added to the total count of $m$ sets satisfies $m*Lap(1/ \varepsilon)$. Although the responses to batch queries on a large number of small sets can be used to calculate data distribution over different attributes or different value divisions of attributes, the transformed distribution could be less accurate than the results from multiple \highlight{data requests} using coarse granularity set conditions but with a small privacy budget respectively. \highlight{In summary, data exploration needs to coordinate data requests by overall planning multiple set division schemes.}

\subsection{Use Case} 
\label{sec:uc}
Lucy is a medical analyst who wants to identify which living habits can decrease the risks of type 2 diabetes. She has successfully applied for a privacy budget to access \highlight{a questionnaire dataset, including descriptions of citizens' living habits (e.g., sleeping hours, eating habits, fitness situations) and their health status.} Due to her limited privacy budget, Lucy has to focus on the attributes that are more likely to reflect the potential prevention of the disease and bargain with the mechanism of \highlight{DP to get more accurate data distributions as convincing providence.}

Lucy first prioritizes the attributes of \textsf{coffee intake} and \textsf{taste preferences}. The distribution of diabetic conditions over either both of the attributes or each of them could be useful. Lucy can divide all individuals in the database into small sets according to the value pair of the two attributes and spends a partial budget to implement batch queries on the total amount of patients in each set. As for the distribution over a single attribute, Lucy can get a relatively inaccurate distribution by adding the requested results on small sets because the sets corresponding to value divisions of a single attribute are combinations of those small sets. Lucy can also spend more privacy budget to request distributions over single attributes. To judge whether the extra budget is necessary, Lucy needs to assess the effect of noise and learn about the accuracy of the calculated distributions. If the accuracy is acceptable, \highlight{Lucy can spend the remaining privacy budget to request and explore correlations between diabetic conditions and other attributes, like \textsf{fitness situations}.}

\subsection{Requirement Analysis}
\label{sec:req}

Privacy restrictions have two primary impacts on the visual exploration processes~\cite{zhang2020investigating, wang2017utility}: (i) users have to spend more cognitive energy when formulating a data request, and (ii) each exploration result contains uncertainty caused by noise. To understand how a visualization-driven approach could help mitigate these impacts, we conducted a pre-study with two experts \highlight{(E$_1$ and E$_2$)} in \highlight{DP respectively}, where each expert had at least three years of research experience. \highlight{In the study, we discussed the two impacts with each expert, recorded their opinions, and refined specific requirements derived from each impact.} 

\textbf{R1: Get advice on data exploration.}
\highlight{When deciding on the next step of data exploration, multiple sophisticated factors need to be considered due to the limitation of DP} and autonomy in data exploration. Users could take advantage of suggestions from the models.

\begin{itemize}
    \item \textbf{R1.1: Sketch the exploration intent.} To come up with effective suggestions on data requests, the recommendation model should save the privacy budget on what users attach importance to. The recommendation model needs to fully understand and follow the exploration intention of users. 

    \item \textbf{R1.2: Express prior knowledge.} \highlight{The effect of uncertainty depends on the requested values. E$_1$ told us that response simulation based on prior knowledge can embody the impact of uncertainty and therefore help the understanding of uncertainty.} The prior knowledge in the scenarios of data exploration includes the public distribution of non-sensitive data and inference results based on domain knowledge from users. Either of them can contribute to the decision on how many sets should be split to request statistical results---if the values of an attribute have a jitter distribution, a finer division could better reveal the distribution details. 

    \item \textbf{R1.3: Browse feasible set division for data \highlight{exploration}.} There could be various choices of set division for batch queries because the attributes of interest have multiple combinations and values of attributes can be divided by different segments. Data requests employing different set divisions feedback distinct data patterns and exert inconsistent effects on the consequent exploration process. \highlight{E$_1$ reckoned that set division schemes need graphic illustrations to facilitate sorting out exploration processes.}
    
\end{itemize}

\textbf{R2: Budget with the understanding of noise.}
The size of noise in \highlight{DP} is controlled not only by the privacy budget paid by users but also by randomness. Users have to determine how much of a privacy budget to set based on the likelihood of noise and the underlying impact.

\begin{itemize}
    \item \textbf{R2.1: Preview the noise effect before the budget is spent.} \highlight{DP} compels data requests to be irrevocable. \highlight{E$_2$ mentioned that users need to confirm the effect of noise caused by DP is acceptable and affordable by previewing the effect of noise controlled by a privacy budget.} 

    \item \textbf{R2.2: Learn about the existence of noise in the actual feedback.} Note that the simulation results for previewing cannot be implemented on the data records in the databases unless privacy budgets are paid. There could be differences between the preview version and the actual feedback. Users still need to learn about the latter. \highlight{Also, E$_2$ told us that users may ignore the uncertainty when browsing a distribution described by a group of specific values. A feasible solution provided by E$_2$ is to remind users of uncertainty through the concept of confidence intervals.}

    \item \textbf{R2.3: Understand the random noise with instances.} \highlight{E$_1$ suggested disclosing the mechanism of noise by showing possibilities, which is similar to instance-based interpretation~\cite{kay2023ggdist, nanayakkara2022visualizing}.} 

\end{itemize}

\begin{figure*}[!ht]
    \centering
    \includegraphics[width=.85\textwidth]{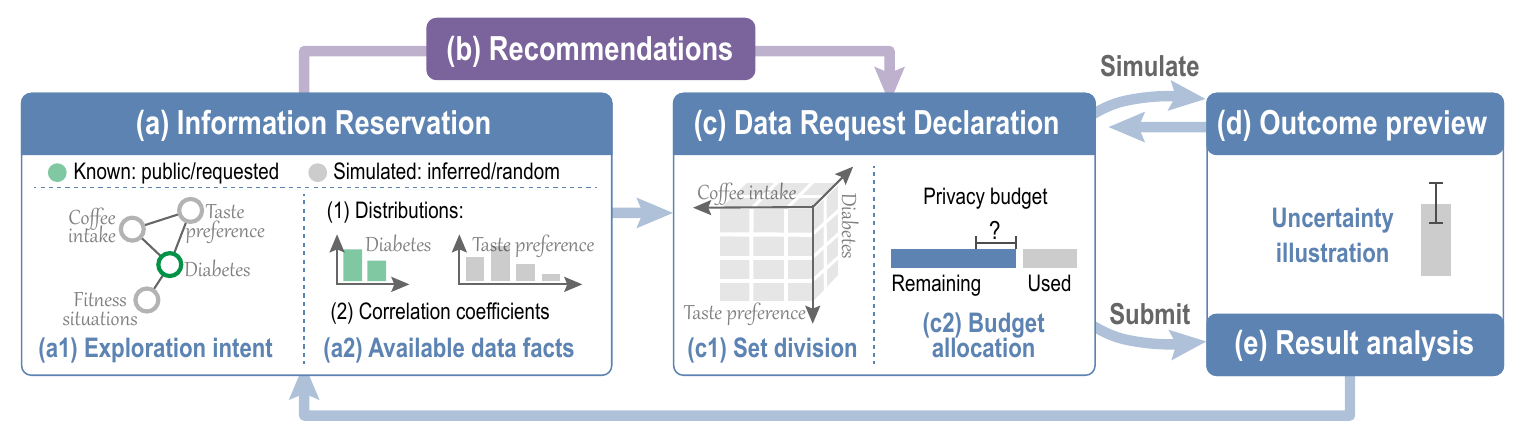}
    \caption{The \highlight{pipeline} for visual exploration with the restriction of \highlight{DP, which mainly consists of a recommendation model and three iterative modules.} In the information reservation module, we take the use case in Sect.~\ref{sec:uc} as an example.}
    \label{fig:pip}
\end{figure*}
\section{Approach}
In this section, we introduce the proposed \highlight{pipeline} for the visual exploration of datasets protected by \highlight{DP} and modules.

\subsection{Pipeline}
\label{sec:ppl}
To satisfy the requirements mentioned in Sect.~\ref{sec:req}, we refined the \highlight{pipeline} into several steps, as shown in Fig.~\ref{fig:pip}. \highlight{Following our pipeline, users can determine exploration strategies and understand the uncertain effects on exploration results when the data interface is guarded by differential privacy.} Seeking effective recommendations for exploration strategies (\textbf{R1}), users are first invited to reserve information about their exploration intent (\textbf{R1.1}\highlight{, Fig.~\ref{fig:pip}(a1)}) or inference on data facts related to the exploration intent (if any, \textbf{R1.2}\highlight{, Fig.~\ref{fig:pip}(a2)}). The exploration intent can be described as distributions of interest, which could be the distributions of a single attribute describing data overview or joint distributions of multiple attributes demonstrating attribute correlations. The above input will be taken into account by the recommendation model \highlight{(Fig.~\ref{fig:pip}(b))} to yield suggestions. Next, users need to declare a data request, consisting of a set division \highlight{(Fig.~\ref{fig:pip}(c1))} and a privacy budget \highlight{(Fig.~\ref{fig:pip}(c2))}, which is no more than the remaining, to drive data exploration. Before making a submission, users can browse suggested data requests (\textbf{R1.3}) and test each of them by simulating the data request on simulation data constructed by all known data facts (\textbf{R2.1}\highlight{, Fig.~\ref{fig:pip}(d)}). After submitting a data request, users can examine the feedback (\textbf{R2.2}\highlight{, Fig.~\ref{fig:pip}(e)}). Aimed at explaining the randomness of noise, users can further browse possible distributions (\textbf{R2.3}) of the actual data calculated by randomly removing a noise. The feedback will be integrated into the known data facts to support the subsequent data exploration, which can also be reviewed by users.

\subsection{Recommendation of Exploration Strategies}
\label{sec:rec}
We employ a recommendation model to suggest exploration strategies. \highlight{An exploration strategy refers to a sequence of data requests that can coordinate to achieve all attribute distributions and attribute correlations specified as the exploration intent. Variables in a data request consist of a set division scheme (i.e., a set of attributes and corresponding value divisions), the order of the request in the request sequence, and a privacy budget. To focus on effective strategies, we first enumerate strategy \textit{prototypes} by identifying groups of attribute sets that can generate feasible groups of set division schemes. For this goal, we describe multiple data distributions by a graph model, where nodes denote distributions of single attributes and edges denote correlation distribution of two attributes. The graph model constructs an intent graph according to the exploration intent reserved by users (Fig.~\ref{fig:pip}(a1)). Similarly, distribution information returned by a data request employing a grid partition (see Fig.~\ref{fig:sd}) can be described as a complete graph with all nodes in the attribute set employed by the set division scheme. Then, the enumeration problem can be solved by listing groups of complete graphs that cover all edges and nodes in the intent graph.}


Next, we \highlight{complete each prototype and generate strategy candidates. Considering that data exploration scenarios lack exploration demonstration and labeled data, we employ a reinforcement learning model} trained by Q-learning~\cite{zhang2024convergence}. Specifically, we describe an exploration strategy as making a series of actions. We explain the definition of actions and the incentive mechanism for an action as follows.

\textbf{Action space:} Each action describes a data request with a privacy budget to be paid and an employed set division. The budget number has to be less than the remaining privacy budget. \highlight{The set division can be an unused attribute set from the prototype integrating with value division schemes for each attribute.} 

\textbf{Incentive mechanism:} We quantify the bonus with three considerations: (i) users prefer accurate illustration for what they have listed as exploration intent, (ii) hubs in the intent graph may correspond to the focus of the exploration intent, and (iii) the saved budget can support more data exploration. \highlight{Following the first consideration,} we identify the joint distribution represented by each edge in the intent graph as \highlight{an object for accuracy calculation.} The bonus for each data request further weights the accuracy of each target distribution by the average betweenness centrality of the two nodes. \highlight{Finally, we assign an extra bonus for budget savings,} a one-time reward that is only issued for the last action in each strategy candidate. Implementation details of accuracy calculation, data simulation for accuracy estimation under the privacy limitation, and calculation of the bonus for budget consumption can be found in the following paragraphs.

\textbf{Accuracy calculation:} We quantify accuracy according to two types of relative errors, which are structural errors caused by inconsistent set division and numerical errors led by noise. In our scenario of data exploration, data distribution can be described by individual numbers in discrete sets. Statistical results in a coarser granularity can provide fewer details of a distribution, which leads to structural errors. The ideal granularity for descriptions of a distribution is the finest granularity allowed by the data interface. Nevertheless, the benefits from details may not linearly increase with the decrease of granularity sizes~\cite{sahann2021histogram}. To quantify the structural error for a set division, we first generate two descriptions of the target distribution: an ideal description providing the \textit{actual number} of data records in each grid specified by the ideal granularity, and the requested description providing the \textit{requested number} of data records in each grid defined by the given set division. Next, we discretize the requested descriptions to align with the former grid and yield corresponding \textit{reconstructed numbers}. There exists a structural difference between each pair of the actual number and the reconstructed number. We quantify the structural error $E_s$ by calculating the average of the relative error (i.e., the difference relative to the actual number) in each grid. The noise generated by \highlight{DP} also adds numerical error to the requested result. The confidential interval of a requested value depends on the paid privacy budget. Nevertheless, the target joint distribution could be indirectly described by the requested values---a data request can employ more than two attributes to define the set division and recover the target distribution by sum operations. In this case, the privacy budget could exert a higher influence on the reconstructed description. Like the structural error, we also consider the number of data records in each grid split by the ideal granularity as the target number. We further calculate the $95\%$ confidential interval of the target number. The half length of the interval relative to the actual number is considered as the numerical error $E_n$. The sum of the structural error $E_s$ and the numerical error $E_n$ is employed as a penalty, namely a negative bonus, for accuracy assessment. 

\textbf{Generation of simulated data:} The disclosure of the original data is forbidden until the privacy budget is paid. In practice, we generate non-public descriptions of sensitive distributions for accuracy calculation based on simulated data. \highlight{Specifically,} we apply a Gaussian copula \highlight{to simulate the multi-variable probability distribution of records based on the input and} generate $N$ samples, where $N$ is the number of records in the database. \highlight{Available input consists of} distributions over each attribute and Spearman's rank correlation coefficient between each pair of non-categorical attributes. 
Besides public information, the input of the Gaussian copula could be derived from feedback on previous data requests, the prior knowledge of users, and random results (set as default). The reason why we employ random results instead of uniform distribution is because the latter yields consistent simulations, which may lead users to ignore diverse possibilities.

\textbf{Bonus for budget consumption:} Ideally, budget consumption should be sufficient and consistent with the progress of data exploration, otherwise the remaining budget could either fail to enable further data exploration or be wasted. The progress of data exploration can be estimated by users. Nevertheless, the estimated progress rate could be inaccurate, especially at the beginning of the exploration process. 
\highlight{Thus, we simulate a growing expectation of consistency in the exploration process.} The bonus for budget consumption $B_{bc}$ is quantified by the following equation:
$$
    B_{bc}=w*N*\frac{p+1}{2}*|p - \frac{\varepsilon_{total}-\varepsilon_{remain}}{\varepsilon_{total}}| ,
$$
where $\varepsilon_{remain}$ and $\varepsilon_{total}$ are the remaining privacy budget after the exploration strategy is implemented and the total privacy budget, $p$ is the rate of exploration progress estimated by users, $N$ is the number of records in the database, and $w$ is an empirical parameter (default as $-0.1$) used for bonus weighting. We default $p$ as 100\%, which indicates that it is feasible to spend all remaining privacy budget to satisfy the specified exploration intent. 

\subsection{Interface Design}
\highlight{Defogger's interface is designed to support the pipeline (Sect.~\ref{sec:ppl}) and refined based on feedback provided by one participant in a pilot user study.}
As shown in Fig.~\ref{fig:teaser}, our interface includes three coordinated views, which display the information reserved by users, recommendations for data request declaration, and illustrations for uncertainty caused by noise.

\subsubsection{Information Reservation}
The information reservation view (Fig.~\ref{fig:teaser}(a)), represents the progress of data exploration by visualizing the exploration intent of users and the known data facts \highlight{(Fig.~\ref{fig:pip}(a))}. Visualizations in this view are dynamically updated with the exploration progress. As shown in Fig.~\ref{fig:teaser}(a1), data facts specified as the target of data exploration are summarized as an intent graph, where nodes are attributes of interest and edges correspond to correlations among attributes. The stroke color of nodes enables differentiating sensitive attributes and public attributes whose distribution can be checked without spending a privacy budget. The visualizations on nodes demonstrate distributions over the corresponding attributes. The distributions of public attributes are known data facts shown by dark gray lines. The distributions shown in light gray lines are simulated distributions, whose data facts are unknown and waiting for exploration. After the distribution is returned by a data request, a dark gray line that describes the requested distribution will be superimposed on the node. The width of the new line visualizes the length of the $95\%$ confidential interval of the distribution to highlight the uncertainty in the requested results. The distribution in dark gray will be updated if a requested distribution that is supposed to have a higher accuracy is received. Users can identify the differences between the simulation and the requested results with noise. If a large difference is observed, users may need to adjust their exploration strategy. Data facts about attribute correlations specified as exploration targets could be known or waiting for exploration as well. We employ color encodings of edges to make a distinction. Furthermore, we label edges between two non-categorical attributes with values of Spearman's rank correlation coefficient. Exactly like the distributions shown in the node, the values labeled on an edge could have a simulated version and a requested version.

\textbf{Sketch exploration intent:} Users can drag an attribute of interest from the attribute list to the sketch panel. After an attribute is dropped, the sketch panel creates dashed edges from the dropped attributes to other attributes on the panel. Users can click a dashed edge to admit their interest in the correlation between the attribute pair. All nodes and the solid lines in blue are regarded as data facts to be explored and considered to recommend data requests. 

\textbf{Estimate the progress of exploration:} Users can input a value as the estimated rate of exploration progress \highlight{after the current exploration intent is satisfied}. The value is used to calculate bonus for budget consumption (please refer to Sect.~\ref{sec:rec}). After a privacy budget is spent for a data request, the lower bound of the estimated rate will be updated to $\frac{\varepsilon_{total}-\varepsilon_{remain}}{\varepsilon_{total}}$, where $\varepsilon_{remain}$ and $\varepsilon_{total}$ are the remaining privacy budget and the total privacy budget.

\textbf{Generate data simulations:} Simulations of the distribution over a single attribute or correlation metrics are applied to simulate joint distributions, assess the accuracy of feedback, and recommend data requests. If users are not satisfied with the default simulation, they can randomly reset it or specify it according to their prior knowledge. After clicking a node in the intent graph, users can interact with a histogram under the sketch panel to specify the distribution of the selected attribute (Fig.~\ref{fig:teaser}(a2)). The bins of the histogram are in the minimum granularity for data requests. However, the entire range of all bins could be inconsistent with the actual range of data records in the database because data owners may employ larger ranges to avoid leakage of extreme values. As for correlation coefficients between two numerical attributes, users can input the inferred number by double-clicking the annotation on the edge connecting the corresponding nodes.

\subsubsection{Data Request \highlight{Declaration}}
\highlight{The second view allows users to declare data requests (Fig.~\ref{fig:pip}(c)) in a panel (Fig.~\ref{fig:teaser}(b1)). Below the panel, we show a list of exploration strategies suggested by the recommendation model, as shown in Fig.~\ref{fig:teaser}(b2), to facilitate planning data requests. Each strategy is summarized by an overview glyph that annotates the attribute sets used in each data request on the intent graph by bubbles. The glyph can illustrate how data requests in a recommended strategy are incorporated to achieve the exploration intent. The total privacy budget required by the strategy is marked on the glyph. We also attach details of data requests (i.e., the allocated budget and attribute names) next to the glyph.} 

\textbf{Declare a data request:} Users can \highlight{select a data request in the strategy list to autofill in the panel (Fig.~\ref{fig:teaser}(b1)). Edit operations are also allowed in the panel} before simulating or submitting the data request. 

\subsubsection{Uncertainty Illustration}
\begin{figure}[!ht]
    \centering
    \includegraphics[width=0.99\columnwidth]{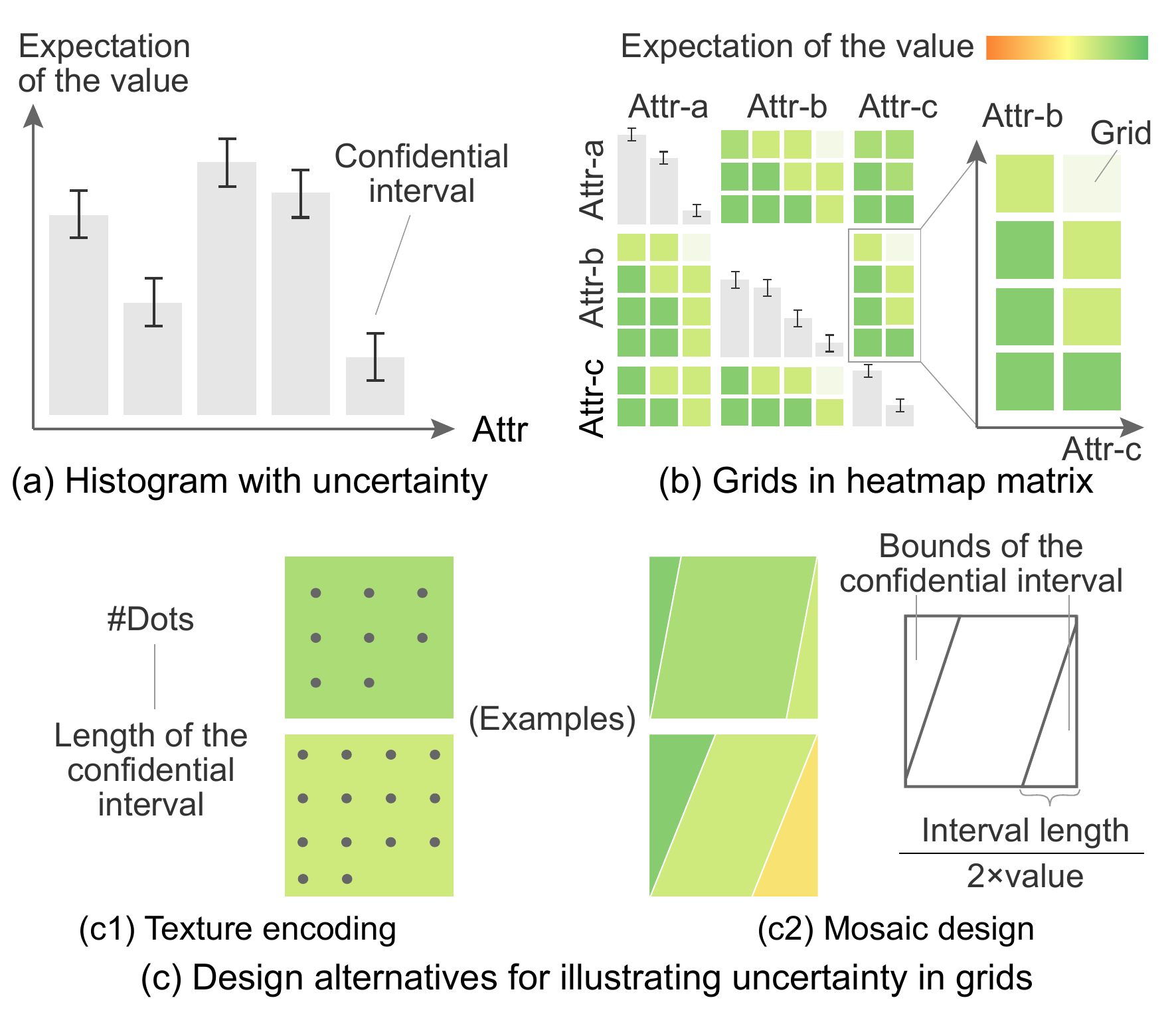}
    \caption{Visualizations for uncertainty illustration. (a) Histogram with error bars for uncertainty in distribution over a single attribute. (b) Heatmap matrix representing correlations among multiple attributes, which requires grid-based uncertainty representation. (c) Two alternatives for uncertainty representation used in heatmap matrices.}
    \label{fig:alt}
\end{figure}

In the uncertainty illustration view (Fig.~\ref{fig:teaser}(c)), users can understand the uncertainty in the responses to data requests \highlight{(Fig.~\ref{fig:pip}(e))} or response simulation by checking descriptions of uncertainty \highlight{(Fig.~\ref{fig:pip}(d))}. To ensure the expressiveness of uncertainty illustration, we integrate visual designs for uncertainty into two relatively universal visualizations for distribution or correlation, which are histograms~\cite{ioannidis2003history} and heatmap matrices~\cite{wang2020conceptexplorer}. To represent uncertainty, we draw error bars on histograms to annotate the range of \%95 confidential intervals, as shown in Fig.~\ref{fig:alt}(a).

Matrices can reflect correlation patterns of multiple attributes (Fig.\ref{fig:alt}(b)). In a matrix, attributes are arranged in a row and a column. Each cell in the matrix displays grids to describe the set-based joint distribution of two attributes. The color of each grid encodes the number of individuals in a set. We use contrasting colors to map a range of values from negative to positive because quantities in grids could be negative due to the noise. Although the correlation between a pair of attributes can be reflected by colorful grids in a cell, common visual representations of uncertainty (e.g., error bars, ribbons) can hardly be integrated into a grid without incurring visual clutter. \highlight{Design schemes that can adapt to grids mainly fall into two categories: overlaid texture and Mosaic design~\cite{goodwin2015visualizing}. The latter represents complicated patterns with multiple colored shapes. We proposed two implementation alternatives (one for each category)}: \highlight{(i)} a texture of dots that demonstrates the size of noise by the density/number of points (Fig.~\ref{fig:alt}(c1)) and \highlight{(ii)} the Mosaic design that divides two triangles from the upper left corner and lower right corner of a grid to visualize the upper and lower bounds of the confidential interval respectively (Fig.~\ref{fig:alt}(c2)). We finally adopted the latter because it is more aesthetic. If the uncertainty in the count encoded by a grid is relatively high, the color differences in the grid will be conspicuous to users. Nevertheless, the color differences are difficult to quantify. To provide intuitive illustrations like dots, we further encode the length of the 95\% confidential interval by the width of the triangle. The width ratio of the triangles to the grid is equal to half the length of the interval divided by the count. \highlight{Further, we employ a rainbow color mapping (instead of a two-color gradient) to reduce the difficulty of distinguishing values, which is mentioned by the participant of our pilot study.} The intent is users will not be bothered by the illustration of the uncertainty if its effect is negligible. For example, the upper example shown in Fig.~\ref{fig:alt}(c2) indicates a smaller effect of noise compared with the lower example. We set a maximum width for the triangles to prevent the middle stripes whose color encodes the requested/simulated numbers from being completely obscured.

\textbf{Filter data records:} As mentioned in Sect.~\ref{sec:ppl}, the sets used in count requests can be defined with multiple attributes. However, even the cells in matrices can only represent the individual number in a set defined by two attributes. To check individual counts in sets defined by more attributes, users can apply value-based filters of the attributes that are used to specify set division in data requests. 

\textbf{Browse possibilities of actual data:} As for uncertainty in the feedback of data requests, users can further browse possible instances of the data distributions inferred by the feedback. After switching to the noise-removed mode,  users can click the button ``Reset'' to generate a new instance of data distribution. The button can be clicked multiple times to support understanding of randomness. The likelihood of each instance presenting coincides with the probability of producing a corresponding amount of noise.

\textbf{Review the summary of requested results:} Users can switch to the summary mode to review all existing requested results in a matrix. If a correlation is requested multiple times, we show the results with the smallest 95\% confidential interval. There could be empty cells in the matrix, which reminds users of exploration gaps.
\section{Evaluation}
\label{sec:eva}
To assess the proposed approach, we carried out a user study and recorded usage processes implemented by two subjects as case studies.

\subsection{Case Studies}
\label{sec:case}
In this section, we outline the exploration process of two of the study participants as a pair of case studies. To support anonymity, we reference these participants with the pseudonyms Tom and Susie.

\subsubsection{Target Clients Identification}
\label{sec:ins}
In this case, Tom played the role of an insurance company employee. His exploration goal was to identify the features of consuming behaviors shared by target clients who are willing to pay high premiums but claim a small reimbursement. For this goal, he needed to explore the client dataset~\cite{case1} with a privacy budget of \highlight{one}. The dataset includes 89,392 client records.

\begin{table}[htb]
\caption{Descriptions of attributes in the database of clients, which were employed in the case. We remark each numerical attribute with the value range and the minimum interval for set division. As for categorical attributes, the remark consists of categorical values. Sensitive attributes are colored in red. (The next table is the same.)}
  \label{tab:insurance}
  \scriptsize
	\centering
\begin{tabular}{l|p{0.4\columnwidth}|p{0.26\columnwidth}}
\toprule
Attribute & Description & Remark\\
\midrule
\rowcolor{red!20}
\textsf{claim\_amount} & Total amount claimed. & $[0,40k], 5k$\\
\textsf{policy} & Active policy of the client. & A, B, C\\
\rowcolor{red!20}
\textsf{cltv} & Client lifetime value.  & $[0,800k], 50k$\\
\bottomrule
\end{tabular}
\end{table}

\begin{figure}[!ht]
    \centering
    \includegraphics[width=0.95\columnwidth]{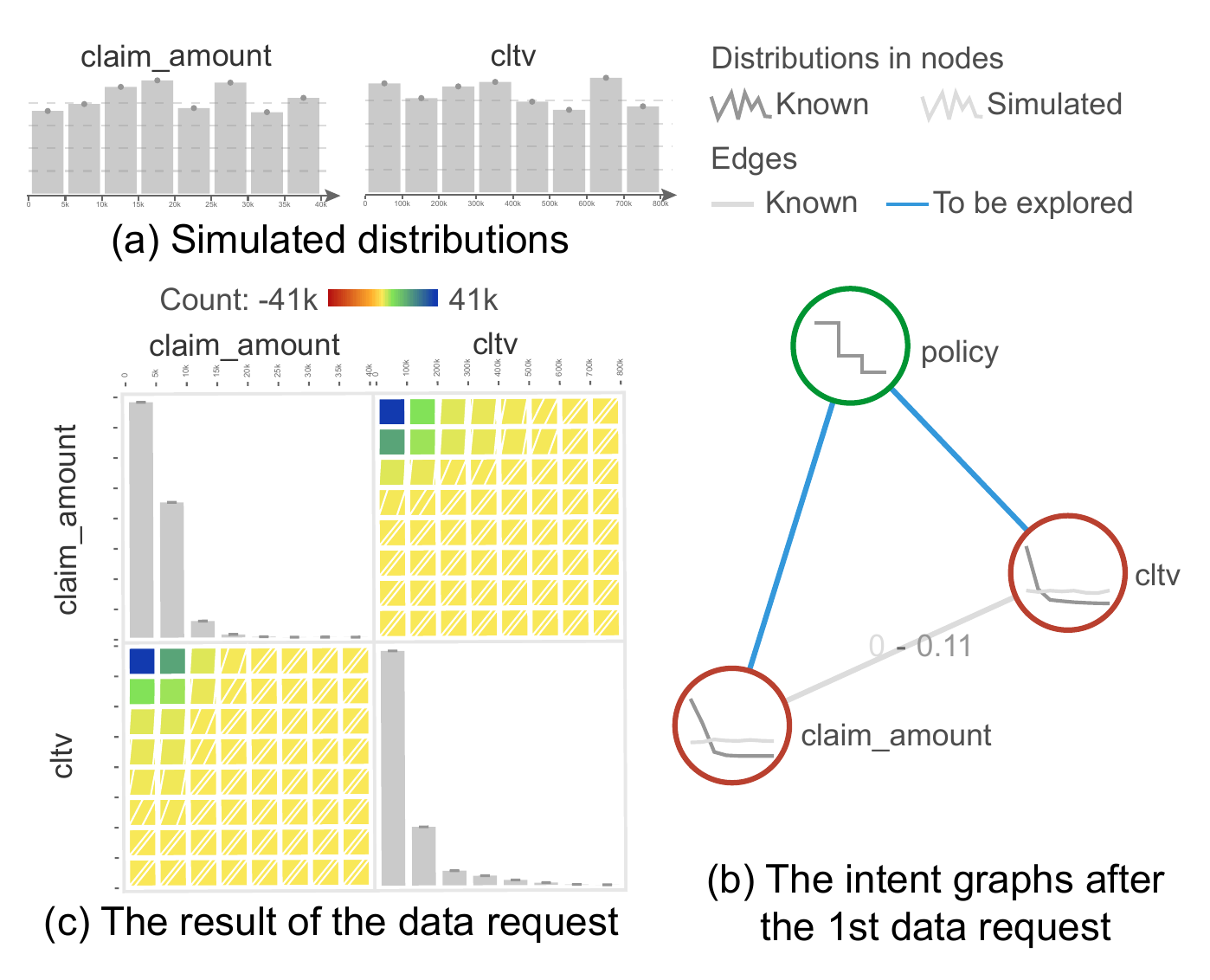}
    \caption{Visualizations used in the first data request.}
    \label{fig:c1-fr}
\end{figure}

The definition of target clients was derived from attributes \textsf{claim\_amount} and \textsf{cltv}, two sensitive and numerical attributes. After checking the descriptions of all 11 attributes (see Tab.\ref{tab:insurance}), Tom guessed that the public categorical attribute \textsf{policy} could be related features of the target clients. He constructed an intent graph with the above three attributes and specified all edges as exploration intent. When browsing the data requests recommended by \techname{} based on the intent graph, he noticed that numerical attributes could contribute to various schemes for set division because of multiple mergeable intervals. Nevertheless, the recommended schemes use the finest granularity directly. Tom found that the recommendation basis of the set division was the simulated distributions, which were generated randomly (see Fig.~\ref{fig:c1-fr}(a)). He reckoned that accurate data simulation is essential for the recommendation model to provide effective suggestions. However, Tom has no prior knowledge of the two sensitive attributes, which can contribute to data simulations. To address this issue, he decided to implement a progressive data exploration, which first requests the joint distribution of the two numerical attributes to find an appropriate set division and then requests statistical results that can describe the correlations among all three attributes.

To start with the progressive exploration, Tom manually declared the data request by specifying the finest grids generated by the two numerical attributes for set division. After weighing the importance of the requested results in the two stages, Tom allocated 20\% of the total privacy budget and submitted the data request. The requested results demonstrated that both the two numerical attributes had a long-tail distribution, which is significantly different from the random simulation (see nodes in Fig.~\ref{fig:c1-fr}(b)). The joint distribution of the two attributes shown in Fig.~\ref{fig:c1-fr}(c) reflected that a majority of individuals gather in the upper-left grids. The size of triangles in other grids indicated that noise added a large amount of uncertainty to the numbers of corresponding sets. However, according to the color of those triangles, the overall distribution pattern could still be observed even when the bounds of confidence intervals are considered.

\begin{figure}[h]
    \centering
    \includegraphics[width=0.99\columnwidth]{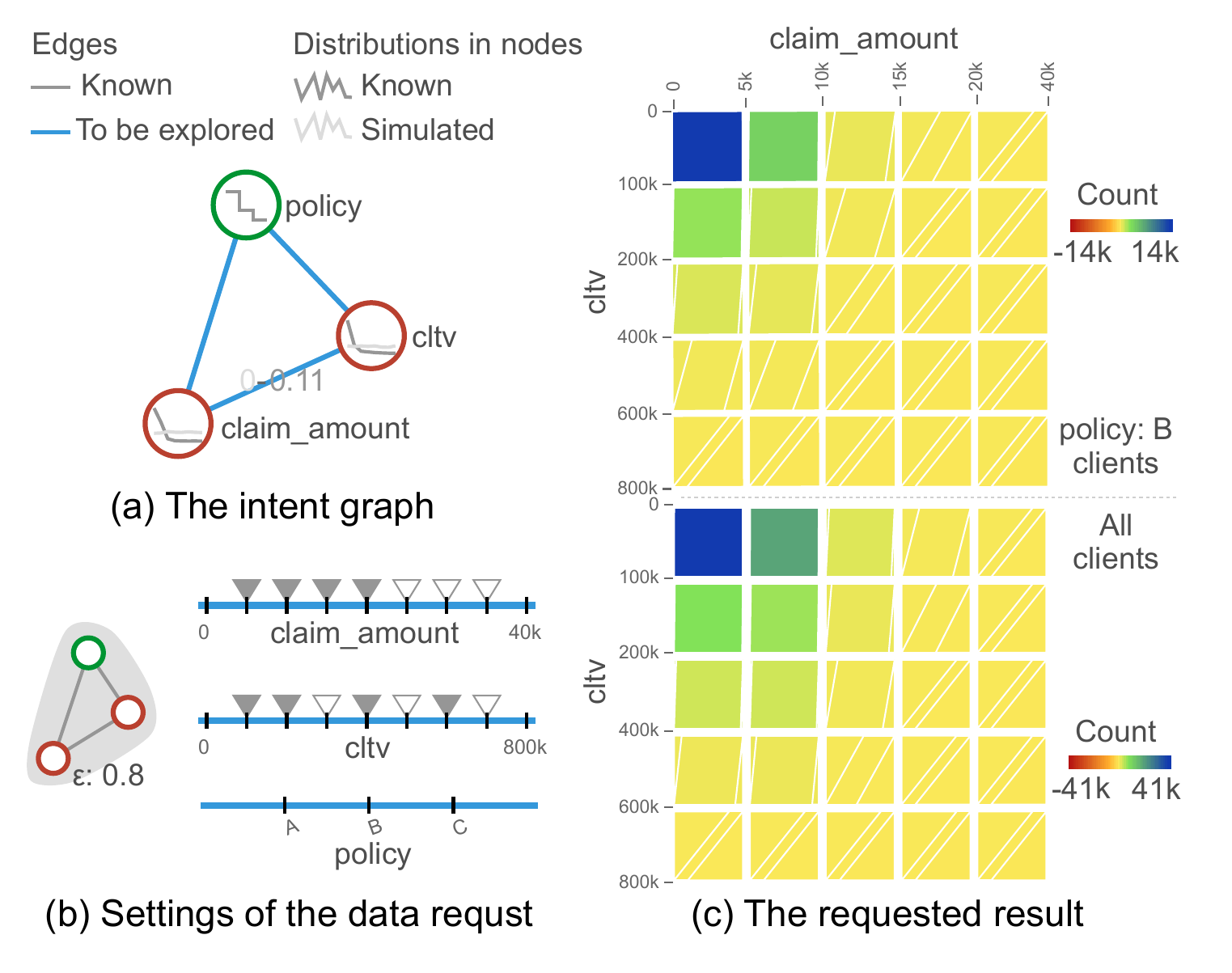}
    \caption{Visualizations used in the second data request, consisting of (a) the intent graph specified by the user, (b) details of request declaration, and (c) distribution comparison between \textsf{policy: B} clients and all clients.}
    \label{fig:c1-sr}
\end{figure}

In the next step, Tom activated the edge between \textsf{cltv} and \textsf{claim\_amount} in the intent graph (see Fig.~\ref{fig:c1-sr}(a)) to complete his exploration intent. \techname{} included the result of the first data request in available data facts for strategy recommendation and updated the recommendation list. To figure out the correlations among all three attributes, Tom selected the recommended strategy that achieved his exploration intent by requesting data with the set division employing the three attributes simultaneously and all the remaining privacy budgets. The new set division canceled the split points with large values of \textsf{claim\_amount}, as shown in Fig.~\ref{fig:c1-sr}(b). Considering that only an extremely small number of clients had claimed high reimbursements, there was no need to make a further distinguishment. Tom believed that merging those intervals would not prevent him from achieving his exploration goals. Thus, he submitted the data request as recommended. The requested result (see Fig.~\ref{fig:c1-sr}(c)) indicated that clients who activated \textsf{policy: B} were more likely to be with a smaller \textsf{claim\_amount} than others with similar \textsf{cltv}. Tom considered his finding reliable because the triangles in upper-left grids are invisible (see Fig~\ref{fig:c1-sr}(b)), which implies that the data pattern has slight uncertainty.

\subsubsection{Demographic Information and Hepatitis B}
\label{sec:hep}
Susie was interested in the correlations between demographic information and the risks of hepatitis B. The second case describes how Susie explored the result of the National Health and Nutrition Examination Survey on 7,824 individuals~\cite{case2} with a total privacy budget of \highlight{two}. 

\begin{table}[h]
\caption{Descriptions of partial demographic attributes in the database of survey results.}
  \label{tab:health}
  \scriptsize
	\centering
\begin{tabular}{l|p{0.55\columnwidth}|p{0.15\columnwidth}}
\toprule
Attribute & Description & Remark\\
\midrule
\rowcolor{red!20}
\textsf{hepatitis\_B} & Have you been diagnosed with hepatitis B? &  Y, N, idk\\
\textsf{family\_c} & \#People in the household. & 1, 2,..., 7+\\
\textsf{children\_c} & \#Children aged $\leq$5 in the household. & 0, 1, 2, 3+\\
\textsf{teenager\_c} & \#Children aged 6-17 in the household. & 0, 1,..., 4+\\
\textsf{elder\_c} & \#Adults aged $\geq$60 in the household. & 0, 1, 2, 3+\\
\bottomrule
\end{tabular}
\end{table}

Although the survey result includes various record descriptions, Susie followed her exploration goal and constructed her graph of exploration intent with the \highlight{five} attributes shown in Tab.~\ref{tab:health}. The \highlight{five} attributes are connected as a star whose center is \textsf{hepatitis\_B}, as shown in Fig.~\ref{fig:teaser}(a1). Next, Susie needed to input data facts for recommendations of exploration strategies. Among the \highlight{five} attributes, only the sensitive attribute \textsf{hepatitis\_B} lacked the data fact of distribution. Susie needed to describe the distribution of \textsf{hepatitis\_B} by inputting the proportion of three categories: \textsf{Y}, \textsf{N}, and \textsf{idk} (i.e., I don't know). According to the disease proportion of hepatitis B reported by the official website of the World Health Organization\footnote{\url{https://www.who.int/news-room/fact-sheets/detail/hepatitis-b}}, that is 10.5\%, Susie divided a small proportion into \textsf{hepatitis\_B: idk} and divided the remaining with a ratio about 1:10 for \textsf{hepatitis\_B: Y} and \textsf{hepatitis\_B: N} respectively\highlight{, as shown in Fig.~\ref{fig:teaser}(a2)}.

In the step of data request declaration, Susie had to determine the attribute groups for set division and the privacy budget to be paid. Susie sought suggestions by browsing the recommendations of exploration strategies (Fig.~\ref{fig:teaser}(b2)) based on the reserved information, as shown in Fig.~\ref{fig:teaser}(b). A data request employing three attributes for set division drew Susie's attention. To inspect how it works, Susie simulated the data request and checked the simulated results in the uncertainty illustration view. As shown in Fig.~\ref{fig:teaser}(c), the size of triangles in each grid indicates that the noise generated by \highlight{DP} could exert diverse effects on the statistical results of different sets due to the uneven distribution. The effect on dark blue grids corresponding to healthy individuals (\textsf{hepatitis\_B: N}) whose family has no children (\textsf{children\_c: 0}) or no elderly (\textsf{elder\_c: 0}) is small enough to be negligible. While grids of hepatitis patients (\textsf{hepatitis\_B: Y}) fail to reflect trustful correlation patterns. Susie hovered over one of those grids to inspect the specific numbers. It turned out that the half length of the confidential interval is greater than the simulated value in a majority of patient sets, which can hardly form a reliable conclusion. Patterns in those grids are necessary for Susie's exploration goal. Susie thought that her privacy budget might be insufficient to support data exploration of so many details. She had to narrow the scope of her exploration to avoid getting no trustful conclusions. After consideration, Susie decided to focus on the correlation between \textsf{hepatitis\_B} and \textsf{family\_c} and issued a data request that employs the two attributes for set division and spends all her privacy budgets.

\begin{figure}[h]
    \centering
    \includegraphics[width=0.9\columnwidth]{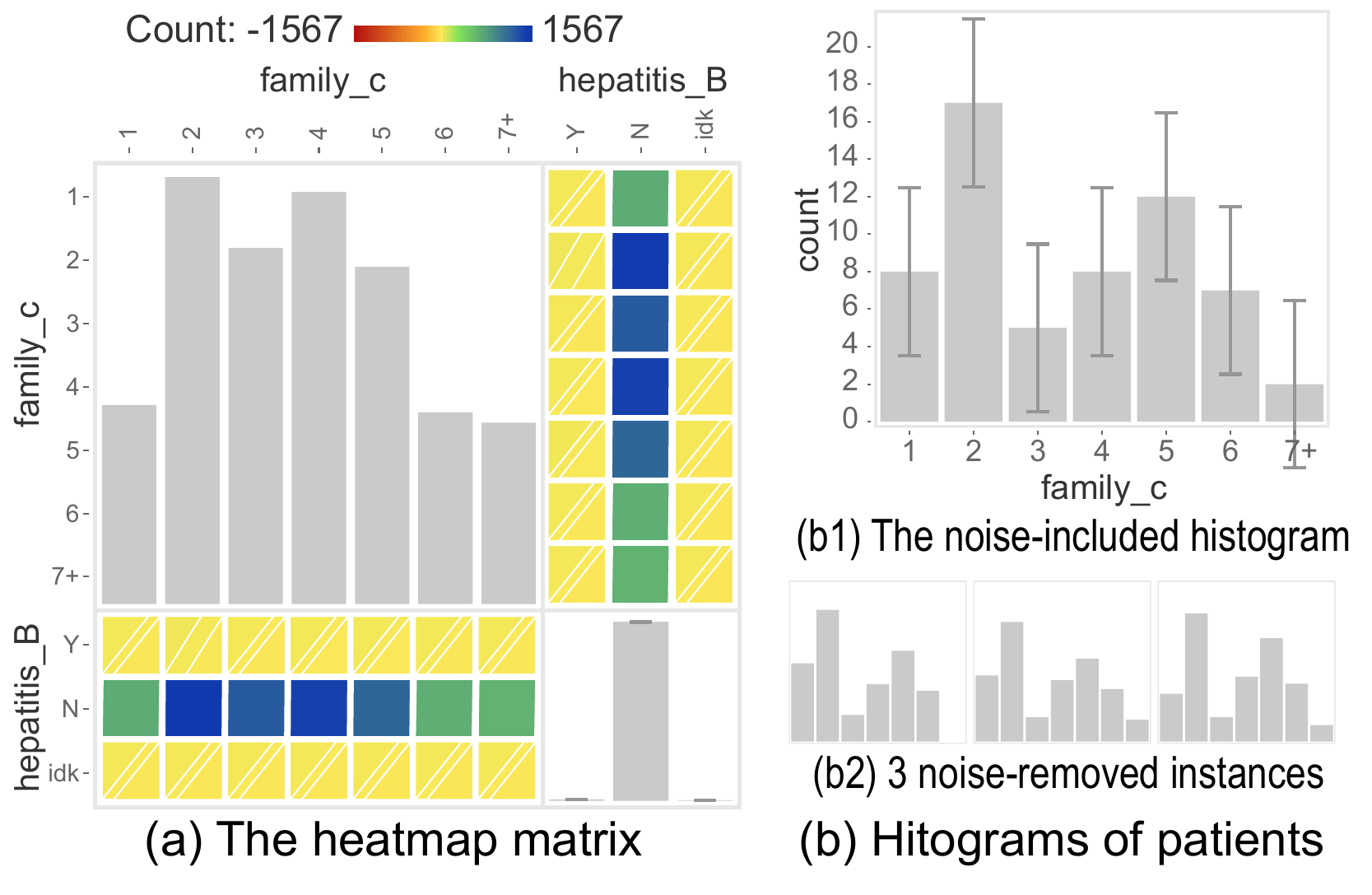}
    \caption{Visualizations for the result of the data request on the correlation between \textsf{family\_c} and \textsf{hepatitis\_B}. (a) The heatmap summarizing the distribution of all individuals in the database. (b) Histograms that demonstrate the \textsf{family\_c} distribution of \textsf{hepatitis\_B: Y} individuals.}
    \label{fig:c2}
\end{figure}

The requested visualization (see the lower-right cell in Fig.~\ref{fig:c2}(a)) demonstrated that the distribution of \textsf{hepatitis\_B} is much more skewed than the simulated data facts. The noise generated by \highlight{DP} leads to more uncertainty in the patterns of hepatitis B patients than Susie expected. Thanks to the all-eggs-in-a-basket move, the data request returned correlation patterns with the smallest uncertainty. To focus on patterns of patients, Susie filtered out \textsf{hepatitis\_B: Y} individuals to compare the \textsf{family\_c} distribution of patients (see Fig.~\ref{fig:c2}(b1)) with the distributions of all individuals (see the upper-left cell in Fig.~\ref{fig:c2}(a)). According to the comparison, Susie came to a preliminary conclusion that individuals who live in a big family (\textsf{family\_c: 3, ..., 7+}) suffer from a lower risk of hepatitis B than those living alone (\textsf{family\_c: 1}). However, the size of the triangles in the row/column of \textsf{hepatitis\_B: Y} demonstrate that the pattern could be not as significant as what is shown in the requested result. To make a verification, Susie further browsed possibilities of actual data by switching to the noise-removed mode. As shown in Fig.~\ref{fig:c2}(b2), there exist differences in the instances of correlation patterns but all instances can contribute to the same conclusion. Finally, Susie confirmed her conclusion and recognized the results of her exploration.

\subsection{User Study}
\label{sec:us}
To evaluate the feasibility of our approach (and the usability of \techname) in supporting \textbf{R1} and \textbf{R2}, we conducted a user study with \highlight{ten} participants (\highlight{eight} graduate students recruited from the university campus and \highlight{two} practitioners in industry recruited from a research group on visual analytics). \highlight{Our subjects have all completed a graduate-level course in data visualization, and are familiar with common interactive visual analysis tools/libraries (e.g., Tableau, D3, R, Python). In other words, they are capable of exploratory data analysis and represent realistic analyst user profiles for \techname{}.} Each subject received \$15 USD for their participation which lasted about an hour.
Because there is not a good competing ``baseline'' tool to compare again \techname, we employed a single approach evaluation study~\cite{tory2013user}.

\textbf{Study Design}: The study comprised three steps. (1) In the training step, we provided an introduction of \techname, consisting of the application scenario, the analysis workflow, the mechanism of the recommendation model, and the interface design based on a demo of \techname. During this, subjects could ask questions at any time and receive detailed answers. The average time for this training step was 22.2 minutes. After training, we showed subjects two databases that could be explored in the next two steps. In addition to descriptions of databases (see Sect.~\ref{sec:case}), we provided a suggested set of exploration goals that subjects could use to help construct realistic exploration scenarios. An example of such a goal is to identify features shared by clients who can bring more benefits to insurance companies. Subjects could also define their own exploration goals of interest. (2) The second step familiarized subjects with the effects of \highlight{DP}, as some did not have experience exploring data within the constraints of \highlight{DP}. In this step, participants needed to select a database. Subjects were told that their trial of \techname{} in the last step had to explore the other database to avoid exceeding privacy budgets. Next, they were allowed to test data requests with the cost of a privacy budget to the selected dataset and check feedback including noise generated by \highlight{DP}, which took about 6.6 minutes on average. (3) In the third step, subjects could freely explore the datasets with \techname{} while employing a think-aloud protocol to catalog their mental processes and actions. This step took an average of 27.7 minutes.
To finish the study, participants completed a questionnaire of 11 usability-based questions (using a \highlight{five}-point Likert scale) and additionally could provide freeform comments/text about the tool and/or its user experience.

\begin{figure}[!ht]
    \centering
    \includegraphics[width=0.85\columnwidth]{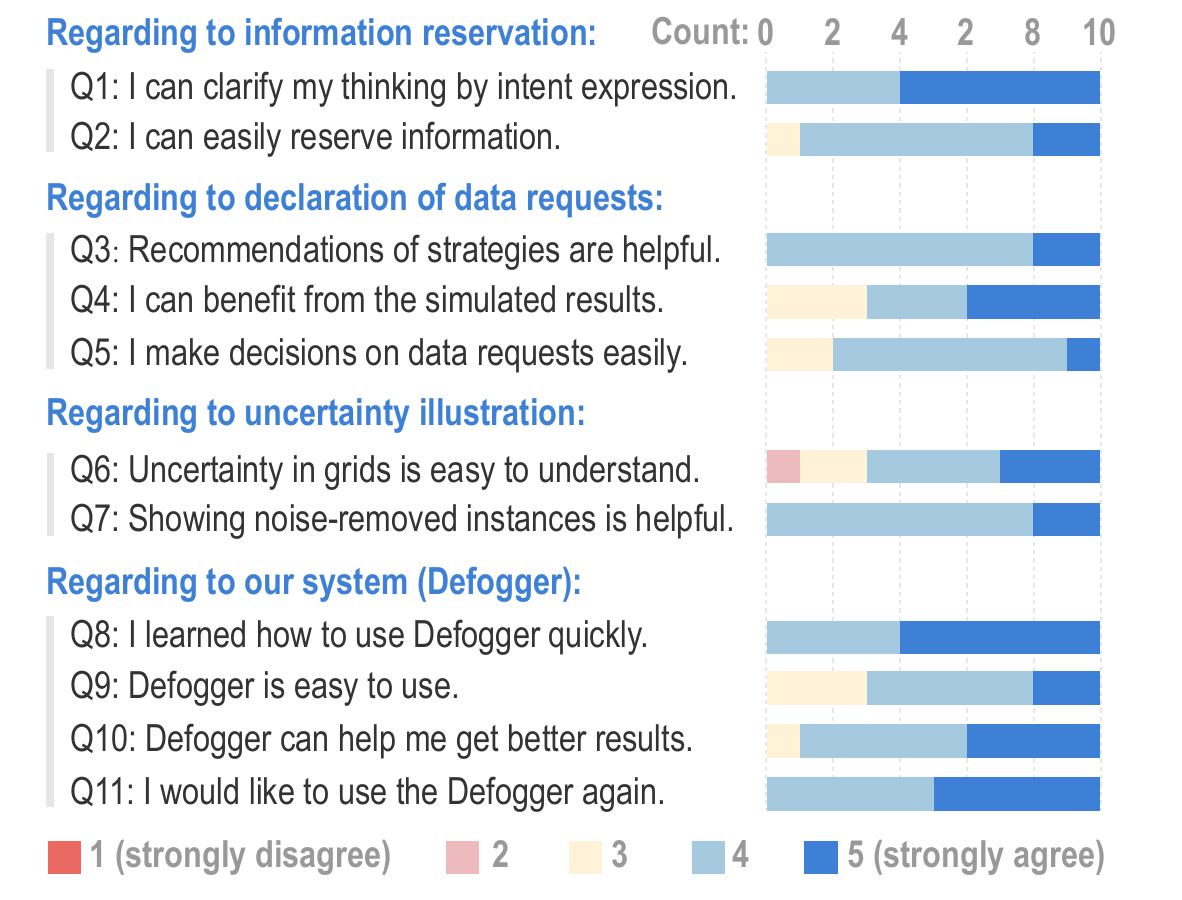}
    \caption{The statistical result of the answers from \highlight{ten} subjects.}
    \label{fig:rs}
\end{figure}

\textbf{Results}: Our questionnaire asked subjects to assess our approach from four aspects, as organized in Fig.~\ref{fig:rs}. We analyze these responses, using freeform comments (both positive and critical) provided by participants for additional context, to assess our approach's strengths and drawbacks.

(Q1--Q2) \textit{Information reservation}: All subjects agreed that sketching the exploration intent allows them to sort out the upcoming exploration process (Q1). As for the input of data facts, a subject expressed concerns about insufficient prior knowledge to specify a reasonable distribution for simulation (Q2). Although two subjects asked for suggestions on distributions, such a request is unattainable because ``free information'' is forbidden by \highlight{DP}. 

(Q3--Q5)
\textit{Declaration of data requests}: Concerns about inaccurate data simulation can cause users to discount the effectiveness of previewing the simulated results (Q4). That said, many subjects still acknowledged the contributions of the preview function to strategy selection because it is necessary to figure out how noisy the requested distribution could be when a certain budget is paid. The recommendations for exploration strategies (Q3), especially budget allocation, were considered valuable suggestions by most subjects. However, one subject complained that making a selection from recommendations was difficult because there were too many choices when the exploration intent included a large number of nodes (Q5). To address this issue, we refined Defogger to sort recommendations according to the assessment result of accuracy (please refer to Sect.~\ref{sec:rec}). Based on the sorted list, users can focus on top strategies when they are not interested in diverse strategy candidates. 

(Q6--Q7)
\textit{Uncertainty illustration}: We received divergent ratings on the representation of uncertainty in grids (Q6). Critical feedback mentioned two drawbacks: unclear information representation and complicated visual design. The subject who rated a 2 for Q6 told us that the visual representation of uncertainty hindered their understanding of the value returned by the data request. This subject showed us a grid overlapped with two triangles that almost took up the entire grid. When it was explained that the returned value is not reliable due to the uncertainty in such a case, the subject approved our design considerations. Other critical comments reflected that the visual design was unintuitive due to the integration of visual elements encoding the confidential intervals. They preferred visualizations representing simple data descriptions, like the heatmap matrices visualizing noise-removal instances (Q7). Those who voted for the uncertainty representation in grids praised the detailed descriptions: ``through the observation of confidence intervals, users can gain a deeper understanding of the uncertainty distribution of the data.'' and ``[i]t is very confusing and requires more time if there is no confidence interval.''

(Q8--Q11)
\textit{\techname{} Usability}: In general, Defogger was considered a handy (Q9) and useful (Q10) tool with a relatively low learning curve (Q8). We receive several positive comments, e.g., ``Defogger provides a lot of auxiliary information that allows me to judge, to a certain extent, how much the queried data has been affected. It allows me to better draw some of the conclusions I want to get.'' One subject especially liked that \techname{} can help users save their privacy budget by avoiding repeated attempts to make unsatisfactory data requests.

\section{Discussion}

To our knowledge, \techname{} represents a first-of-its-kind approach for exploratory visual analysis augmented by privacy-aware recommendation intelligence. 

\textbf{Lessons Learned}:
We briefly summarize three primary lessons learned from the user study and the case studies in Sect.~\ref{sec:eva}.

\textit{Partial users prefer to understand complex information by browsing multiple visualizations demonstrating simple information}: In Sect.~\ref{sec:us}, we discuss how our subjects had opposite views on the Mosaic representation for uncertainty in grids. \highlight{Several subjects who frequently used Python or R as data analysis tools rated the uncertainty illustration view in \techname{} relatively low because they considered that visualizations integrating a variety of visual encodings were not effective at conveying information.} Thus, it is necessary to enable a divide-and-conquer process to learn about complex information. \techname{} enables such a process as an alternative approach to understanding uncertainty by providing noise-removed instances. More broadly, future efforts in this area can investigate how to balance visual complexity with privacy-based situational awareness.

\textit{Experienced data explorers can better steer the exploration process with \highlight{DP} constraints}: The case of target customer identification in Sect.~\ref{sec:ins} describes a smart exploration strategy that is raised by a subject with rich experience in data exploration. In the future, we plan to identify exploration skills from the behavior patterns of experienced explorers and provide users with more assistance by integrating the summarized skills into \techname, as more adept models would likely benefit users at all levels of experience.

\textit{User knowledge should be integrated into the process of strategy customization}: \techname{} recommends exploration strategies based on the exploration intent and available data facts reserved by users. The reserved information is integral to the recommendation model yielding feasible strategies. \techname{} also allows users to declare data requests directly, as users should have the autonomy to formulate their own exploration strategies, especially at the beginning stage when a model will lack a basis to make effective recommendations, or users have difficulties in describing their exploration intent.

\textbf{Current Tool Limitations and Additional Future Directions}:
A positive user experience during visual exploration relies on quick (i.e., interactive) responses to flexible data requests, which relies on robust software engineering and data management. At current, \techname{} focuses on count-based data requests as this is the most versatile type of visual exploration request. However, how to best coordinate data requests of various types in privacy preserving scenarios is a challenging but essential problem for flexible data exploration. When there are a larger number of possible exploration strategies, the employed recommendation model requires a longer period to generate appropriate strategy recommendations. For example, it took the recommendation model about two minutes to generate recommendations for the second exploration intent in the first case study (Sect.~\ref{sec:ins}). While we consider such a duration acceptable for a first-of-its-kind research effort, it is certainly limiting in many real-world scenarios. Feasible future research directions include pre-caching or shrinking the strategy space~\cite{menda2018deep} based on the exploration behaviors of experienced data explorers.
\section{Conclusion}

Visual exploration scenarios that query datasets containing sensitive information must adhere to privacy requirements. In particular, \highlight{DP} and privacy budgets can strain the exploration process, and users need intelligent strategies to optimize their workflows. Based on a pre-study and requirements analysis, we propose and develop a novel visual analysis approach, implemented in a prototype system called \techname, which integrates a reinforcement learning model to recommend exploration strategies based on users' exploration intent and illustrates uncertainty in data distributions by a group of visualizations. We conduct a user study and present two case studies to demonstrate that \techname{} effectively supports users in making valuable findings from sensitive data while still adhering to required \highlight{DP} constraints, and discuss lessons learned that are generalizable to the community and future research efforts. \highlight{The codebase for \techname{} can be found at \url{https://github.com/Vanellope7/Defogger}.}

\section*{Acknowledgment}
This work was supported in part by the NSFC (62202244), ``the Fundamental Research Funds for the Central Universities,'' Nankai University, and by the U.S. National Science Foundation through grant CNS-2224066.

\bibliographystyle{abbrv-doi-hyperref}

\bibliography{template}

\end{CJK}
\end{document}